# Single Photon X-Ray Imaging with Si- and CdTe-Sensors

P. Fischer[a], M. Kouda[b], S. Krimmel[a], H. Krüger[a], M. Lindner[a], M. Löcker[a,*], G. Sato[b], T. Takahashi[b], S.Watanabe[b], N. Wermes[a]

[a] *Physikalisches Institut der Universität Bonn, Nußallee 12, D-53115 Bonn, Germany*
[b] *Institute of Space and Astronautical Science (ISAS) Sagamihara, Kanagawa 229-8510 Japan*

**Abstract**

Studies of a single photon counting hybrid pixel detector for X-ray imaging applications are presented. A silicon- and a CdTe-sensor were bump bonded onto the MPEC pixel readout chip and could be successfully operated. A new USB based readout system was used for data acquisition. Measurements of the performance on the latest MPEC chip and imaging characterization of the sensors are presented.

## Introduction

Single photon counting with hybrid pixel detectors is a method suitable for direct digital imaging [1,2]. The concept of a hybrid pixel detector where sensor and readout chip are separate items allows developing and optimizing the chip and the sensor separately. Furthermore the same readout chip can be operated with different sensor materials.

Hybrid pixel detectors have been first designed for applications in high energy physics [3]. It has been shown that the readout architectures can be modified so that bio-medical applications for the detection of X-rays or ionizing radiation become possible.

The first version of the MPEC-Chip (**M**ulti **P**icture **E**lement **C**ounters) is derived from the prototype pixel chip for the ATLAS experiment at LHC (CERN) [1]. The latest chip generation is the MPEC 2.3 which will be introduced in this paper.

Silicon (Si) is a preferred sensor material for the detection of ionizing radiation because high quality sensors with an excellent detection homogeneity can be made. On the other hand due to its low atomic number Si has a low absorption efficiency for X-rays of with energies of few 10 keV up to 100 keV. This is a serious disadvantage in medical imaging applications where a low patient dose is desired. A material with very high detection efficiency is Cadmium Telluride. CdTe- sensors of only 0.5mm thickness have absorption efficiencies between 90% and 30% for X-ray energies of 40 keV and 100 keV, respectively. Although CdTe is not as well understood as Si as sensor material, progress in CdTe detector development has been recently achieved [4,5]. However, CdTe suffers from charge collection inefficiencies and crystal defects. Furthermore, the bump bonding of CdTe – which is a fragile and expensive material difficult to handle – is still a challenging task. In collaboration with the ISAS a 0.5mm thick CdTe sensor could be bump bonded onto a MPEC2.1 chip and successfully operated [6].

In this paper we report on a MPEC2.3 chip with a CdTe-sensor and a MPEC2.1 with a Si-sensor.

## Detector description

### *The read out chip*

The MPEC 2.1 readout chip is a single photon counting chip with energy windowing [7]. The active area of 6.4 x 6.4mm² is structured into 32 x 32 pixel of 200 x 200 µm² size. A preamplifier, two discriminators and two 18bit counters are integrated into every pixel (fig. 1). A coarse discriminator threshold is set globally and a fine adjustment can be applied dynamically for each pixel. The MPEC 2.3 chip is the latest chip version and is an improvement of the MPEC 2.1 chip. Some crucial parts of the chip layout were reworked, (counter, shift register, multiplexer) and a new window-logic was implemented. Both chips, MPEC2.1 and MPEC 2.3 were processed in the AMS 0.8µm technology. Unfortunately, this process offers only two metal layers which makes an efficient shielding

[*] Corresponding Author, e-mail: loecker@physik.uni-bonn.de

between the digital parts and the sensor and sensitive analog parts difficult.

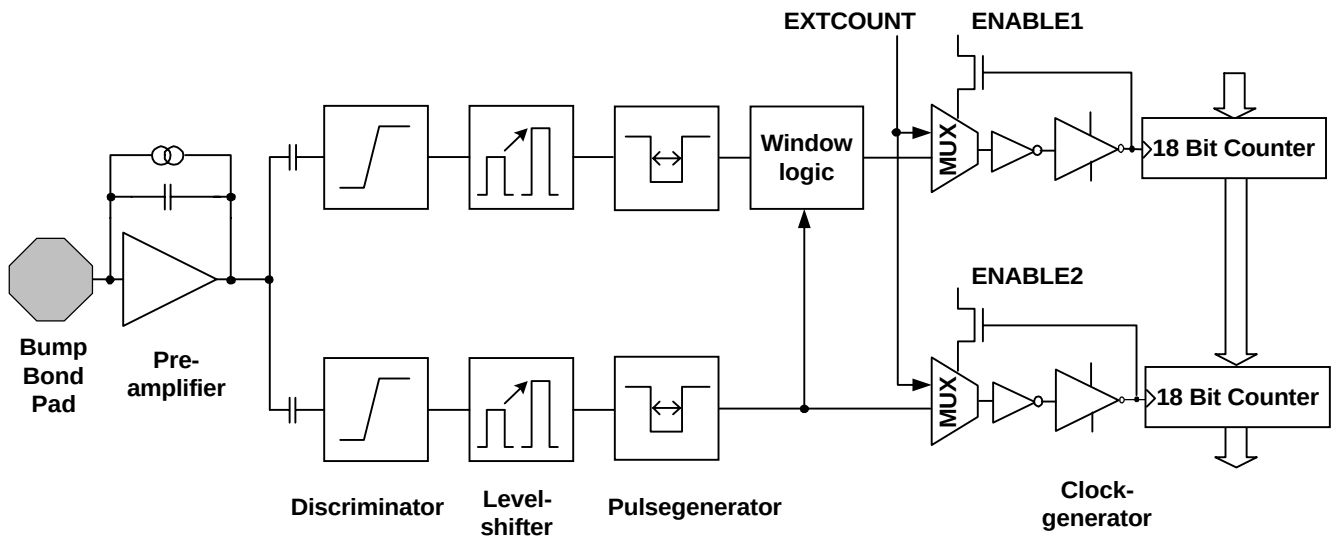

Fig. 1: Schematic of one pixel of the MPEC 2.1 / 2.3 chip

*The sensors*

A n-type Si-sensor of 280µm thickness from CIS was bonded onto the MPEC 2.1 chip by a wafer level solder bump bond process at IZM (Fraunhofer Institut für Zuverlässigkeit und Mikrointegration, Berlin, Germany). The p+ electrode is structured into 32 x 32 pixel, according to the geometry of the readout chip. So far, only a MPEC 1.1 was tested with a Si-sensor [8].

The 0.5mm thick CdTe-sensor was fabricated by ACRORAD and the detector assembling was done by ISAS. Both sides of the ohmic detector are plated with a Pt layer, where again the side connected to the readout chip is structured into 32 x 32 pixel. A special gold stud bump bond technique was used. More details of this sophisticated bonding procedure can be found in [6].

The readout side of both sensors is surrounded by a guard ring of 200 µm width. Currents originating outside the active sensor volume flow into the guard ring and do not cause any artifacts at the edge of the sensitive area.

*The test setup*

All measurements of the MPEC 2.1 chip were done with the testboard system from Silicon Solutions while a newly developed USB-based data acquisition system was utilized for the MPEC 2.3. Besides a faster data transfer this small and handy system is controlled by a laptop computer and can be easily transported and set up. A detailed description will be given in [9].

## Measurements

*MPEC 2.1 / 2.3 characteristics*

The analog part of the MPEC 2.1 and 2.3 chips are almost identical. Table 1 summarizes threshold and noise for the different assemblies. The relatively high threshold is required despite of the low noise because of cross talk from the digital to the analog part via the chip bulk and the sensor. As mentioned earlier a better shielding was not possible with only two available metal layers.

The front end preamplifier and feedback circuit of the MPEC is designed for positive input charges. Therefore, only holes generated in the sensor can be collected. This is not a limitation when operating a Si-sensor where good hole mobility and lifetime are ensured. However, hole trapping and low hole mobility will reduce the imaging performance for the CdTe-sensors.

| | Without sensor | Si-,CdTe-sensor |
|---|---|---|
| Threshold | 2000e | 4000e |
| Noise | 60e | 120e |

Tab. 1: Specifications of the MPEC 2.1 / 2.3 chips

*Thresholds*

An important characteristic for the imaging performance of a photon counting chip is the setting of the thresholds. Both thresholds of a single pixel can be independently adjusted by storing dynamically in each pixel a correction voltage on a capacitor. The unadjusted and adjusted thresholds of the MPEC 2.3 chip are shown in fig. 2. The RMS-values of the threshold dispersion are 180e for the unadjusted case and 10e for the adjusted case (fig. 3).

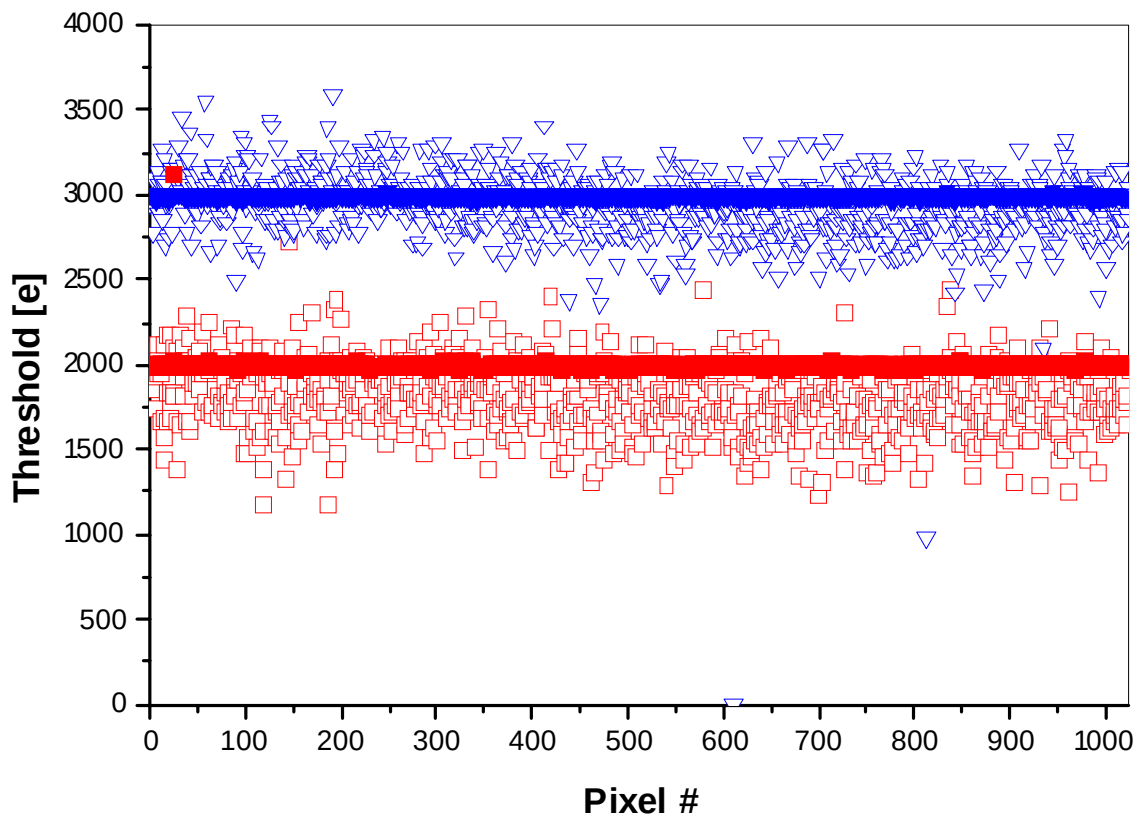

Fig. 2: Upper and lower Thresholds of the MPEC2.3 chip. Open symbols are before, filled symbols are after threshold fine tuning.

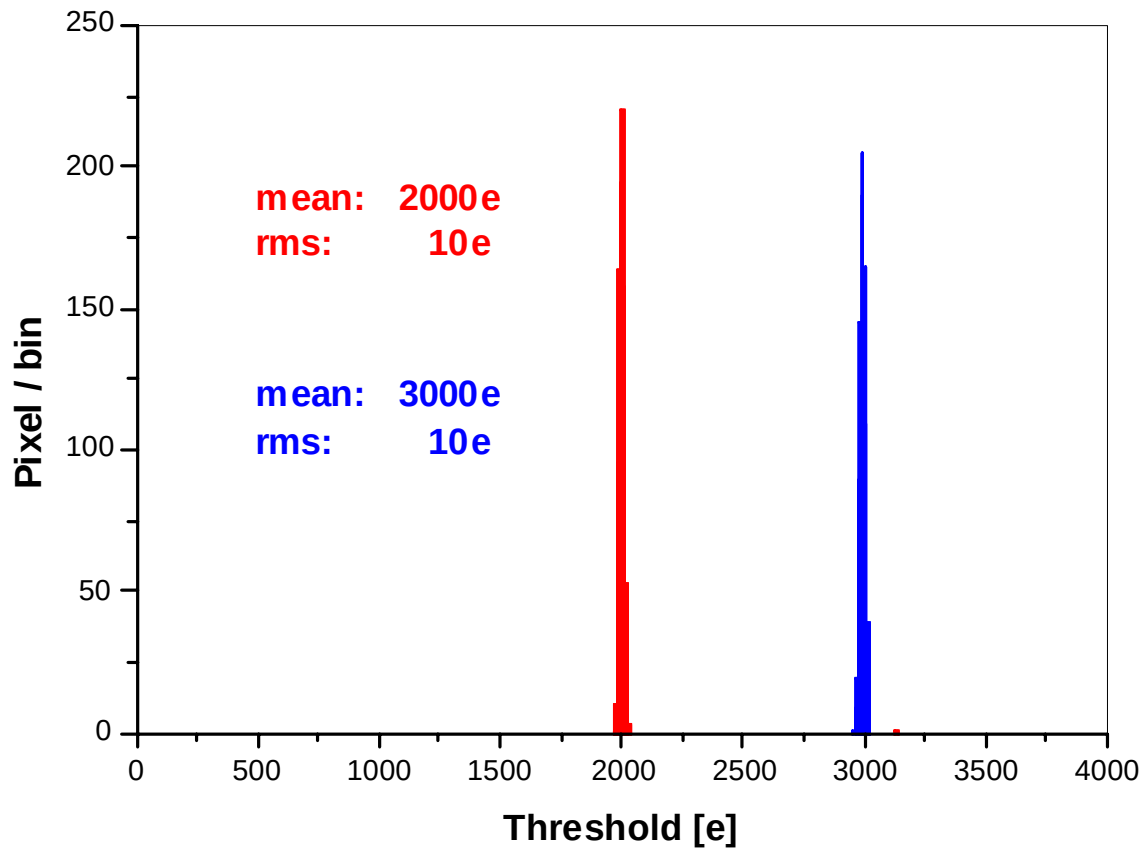

Fig. 3: Dispersion of adjusted lower and upper thresholds of MPEC2.3.

For photon counting with energy windowing the capability of a low threshold dispersion is essential. A low threshold dispersion could already be demonstrated with the MPEC 2.1 in [11]. However, the outer columns of the MPEC 2.1 chip do not work properly, whereas this problem is eliminated in the MPEC 2.3.

As the threshold adjustment is done dynamically the correction voltage needs to be refreshed. In order to minimize the refresh rate, a threshold drift compensation circuit was developed which reduces the drift current from the capacitor [11]. As an example, the measured drift of the lower threshold of the 32 pixels of one column is shown in fig. 4. At the beginning the thresholds were set to 3000±10e. The thresholds were then measured every 2 minutes. The maximum drift rate is only 0.2e/s so that only one fine adjust cycle is needed for a short detector exposure time.

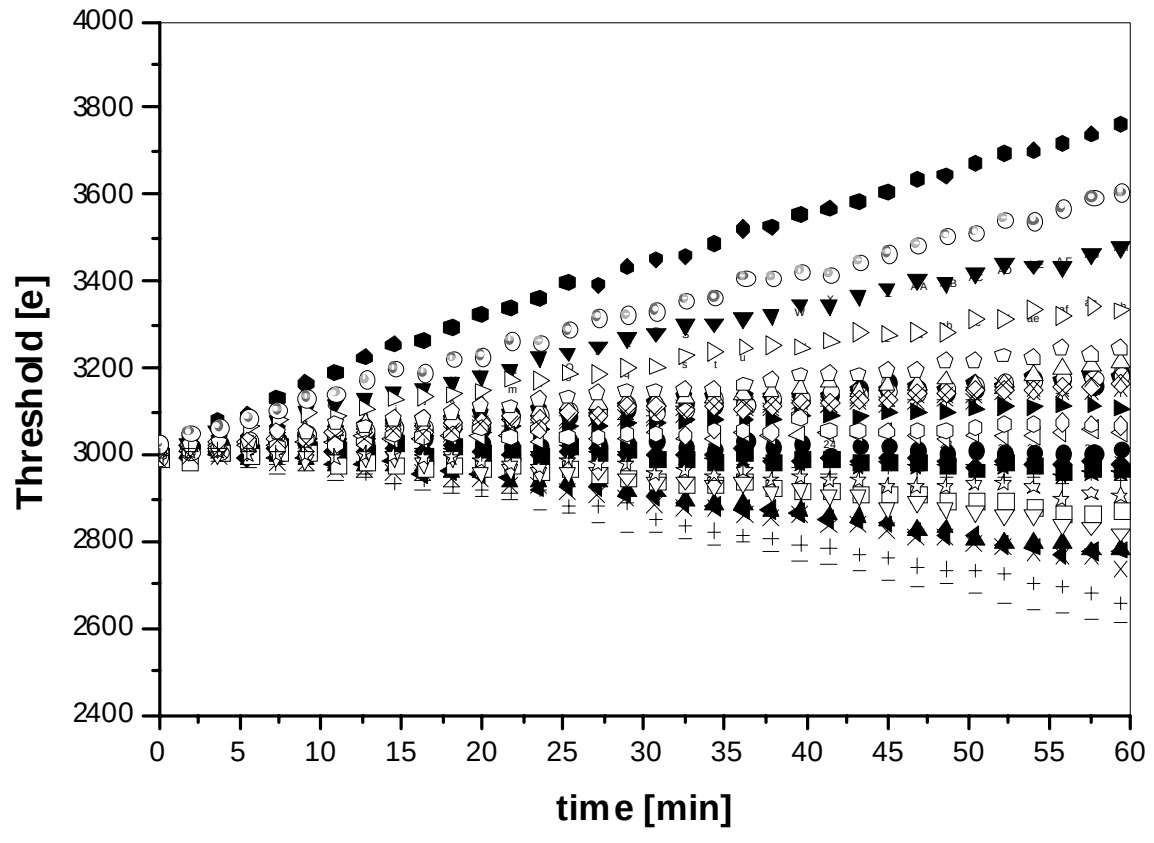

Fig. 4: Treshold drift of 32 pixel within one hour

*Detector leakage currents*

The Si-sensor is a high resistance $p^+$ on n sensor operated in diode configuration and the I-V characteristic is shown in fig. 5.

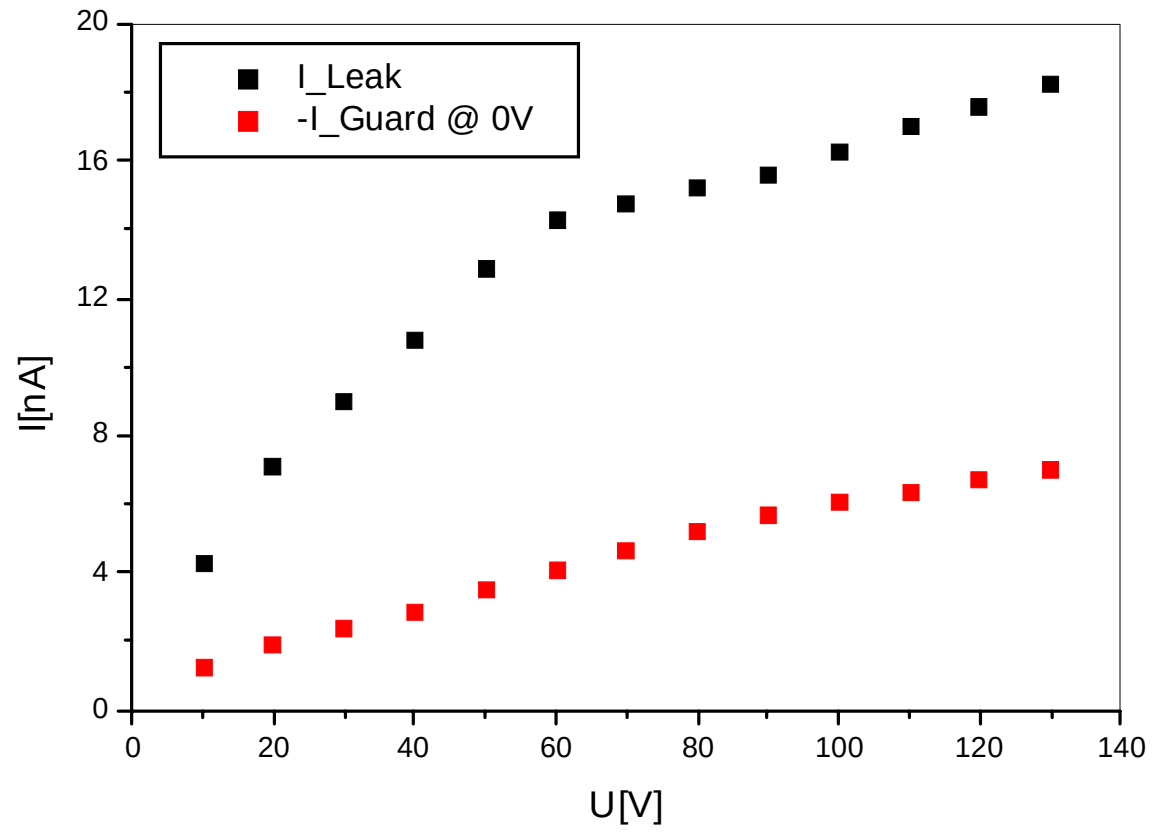

Fig. 5: I-V characteristic of the Si detector

The CdTe detector is operated as an ohmic detector with the Pt/CdTe/Pt electrode configuration. The ohmic I-V characteristic is plotted in fig. 6. A resistivity of about $\sigma = 1 \cdot 10^9\ \Omega$cm is measured.

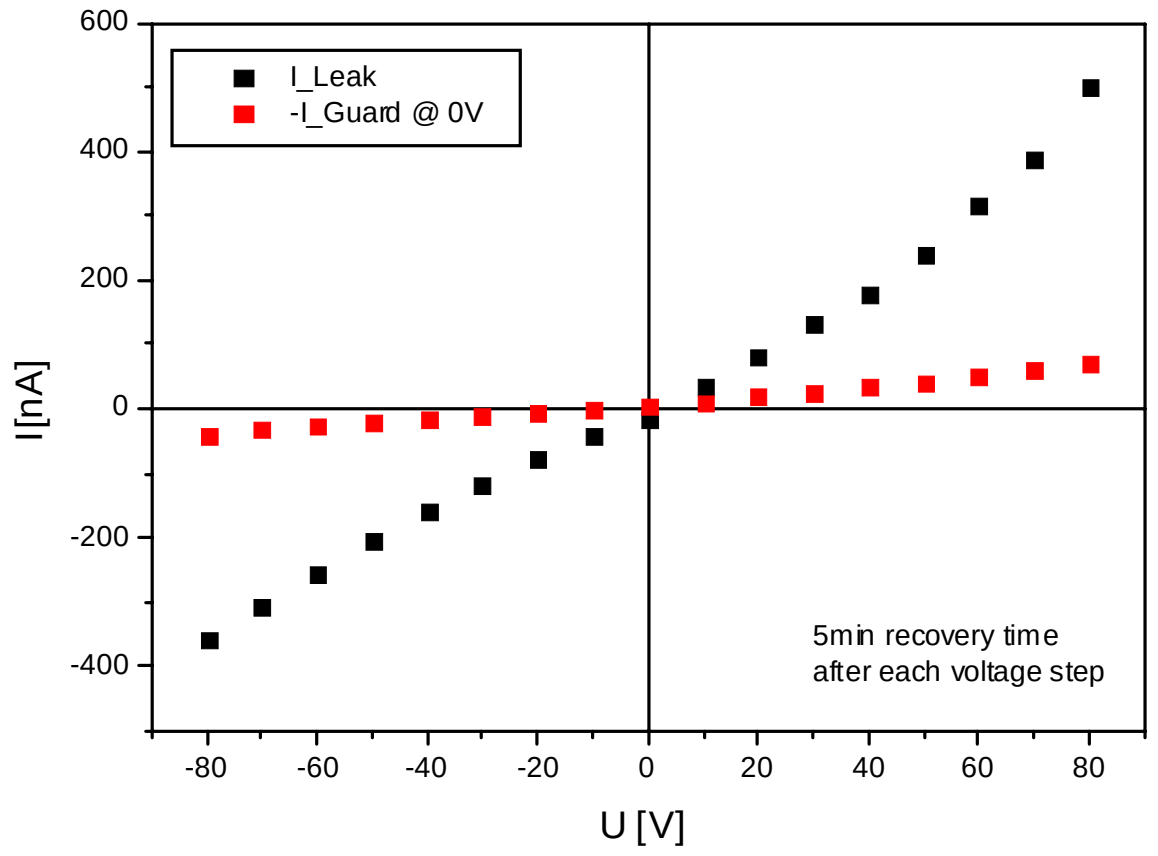

Fig. 6: I-V characteristic of the CdTe detector

## Photon Counting Imaging

*Si-sensor*

After bump bonding of the MPEC 2.1 onto the Si-sensor only one pixel out of 864 did not work correctly, an excellent bump yield which could be achieved for the first time with an MPEC chip. The outer columns cannot be read out properly in the MPEC 2.1 and therefore only 864 pixel out of 1024 can be tested.

A flat field image was taken by exposing the detector during $t_{Si}$=100min uniformly to a $^{241}$Am 60keV photon source. The binned count rates are plotted in fig. 7. The distribution can be approximated by a Gauss-function with mean <n>=822 and $\sigma_{meas}$=53. From this the Signal-to-Noise Ratio $SNR = \langle n \rangle / \sigma_{meas} = 16$ is obtained, a measure used for characterising imaging detectors [11,12].

The radiogram of a screw nut shown in fig. 8 demonstrates the imaging performance of the system.

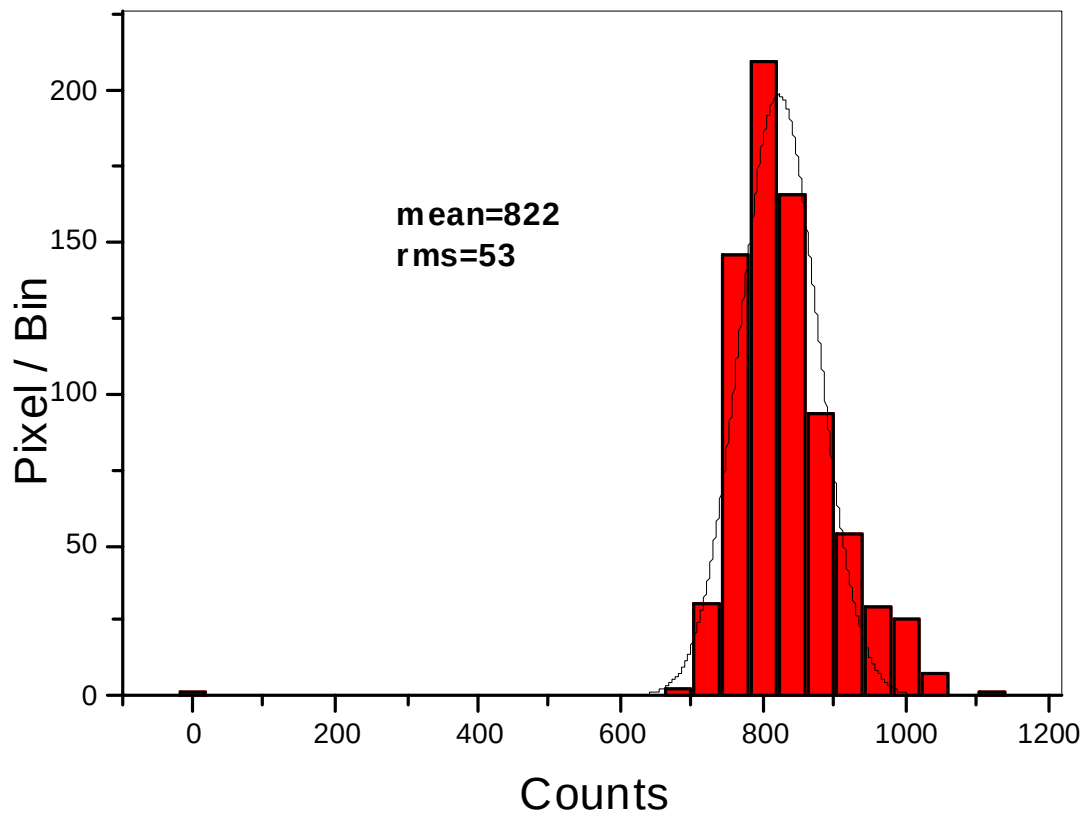

Fig. 7: Countrate distribution of flat field image, Si-sensor exposed to $^{241}$Am 60 keV

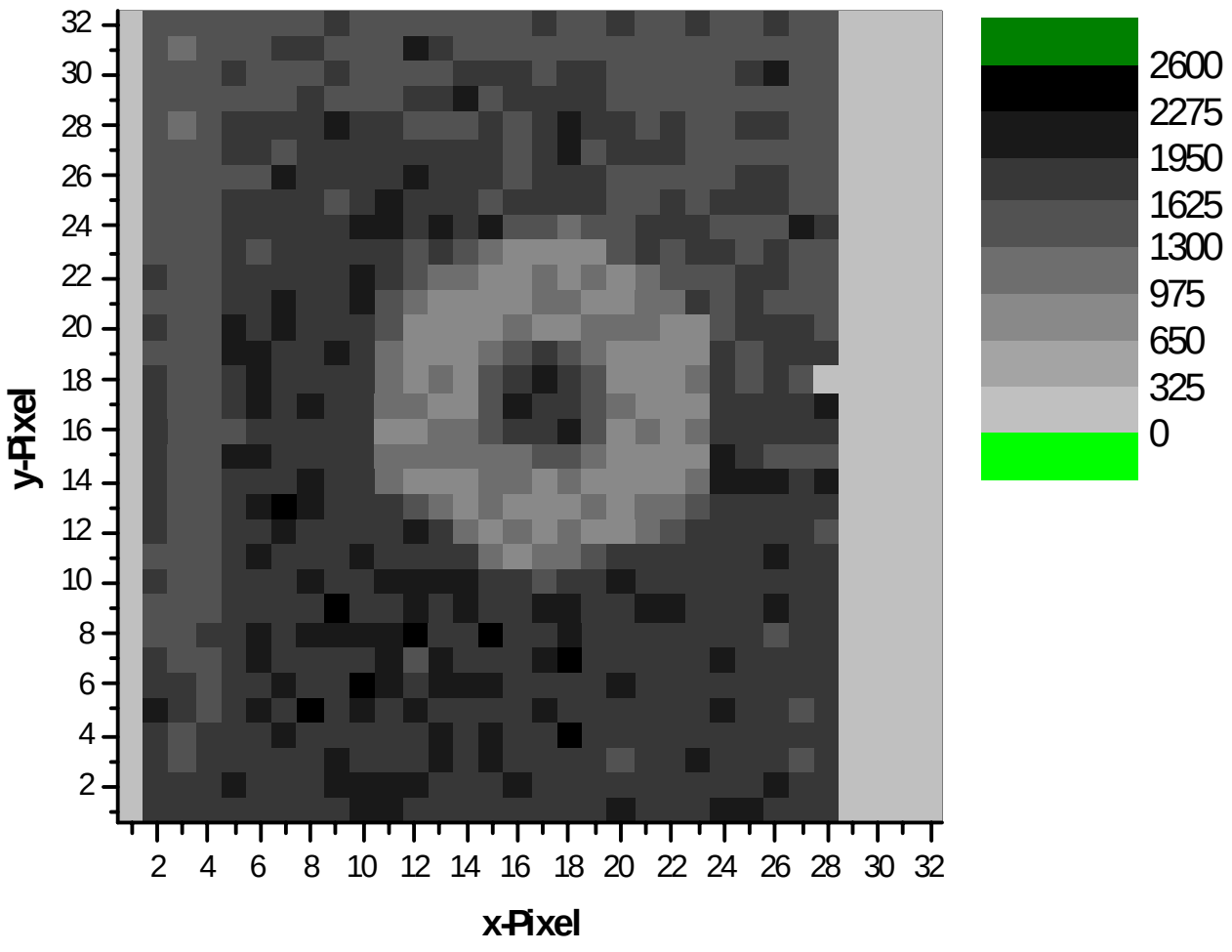

Fig. 8: Shadow image of a screw nut, Si-sensor exposed to $^{241}$Am 60 keV

*CdTe-sensor*

A similar characterization was done for the CdTe detector. Unfortunately after bump bonding the output buffer for four columns was damaged and thus these four columns can not be read out. Regarding all other columns a bump yield of 94% could be achieved, a result similar to [6]. A flat field image with the $^{241}$Am 60keV source was recorded with an identical geometry during an exposure time of only $t_{CdTe}$=1min. With the measured mean <n>=791 and $\sigma_{meas}$=194 a Signal-to-Noise-Ratio SNR=4.1 is obtained. The $\sigma_{meas}$-width of the count rate distribution is caused by statistical quantum noise of the incident photons and the inhomogeneities of the sensor material. The relatively high $\sigma_{meas}$-value for the CdTe sensor is mainly caused by spatial pixel to pixel variations of the charge collection efficiency (CCE). A series of measurements of a *single* pixel has a count rate statistics close to $\sigma_{stat} = \sqrt{<n>}$ as expected from the Poisson statistics for quantum noise of the radioactive source. If a flat field correction is applied to the data set above the count rate spread is reduced to $\sigma_{meas}$=32 and a Signal-to-Noise-Ratio SNR=24.7 is obtained. The flat field correction factors for each pixel $f_{ij}$=<m>/$n_{ij}$ are obtained from one (or more) *independent* flat field images where $n_{ij}$ is the pixel count rate and <m> is the mean count rate of all pixel. Possible contributions for the spatial variation of the CCE might come from the CdTe material or the readout circuit (i.e. shaping time variations).

As expected the mean count rate <n>/t is much higher than for the Si-detector. The ratio of the count rates

$$\frac{<n_{CdTe}>/t_{CdTe}}{<n_{Si}>/t_{Si}} = 96.2$$

corresponds reasonably well to the ratio of the absorption efficiencies [13]:

$$\frac{Efficiency(CdTe, 60keV, 500\mu m)}{Efficiency(Si, 60keV, 280\mu m)} = 99.8\,.$$

A radiogram of a screw nut taken with the CdTe detector is shown by fig. 9 and fig. 10 without and with flat field correction respectively. It can be seen that the increased homogeneity by the flat field correction enhances the image quality significantly.

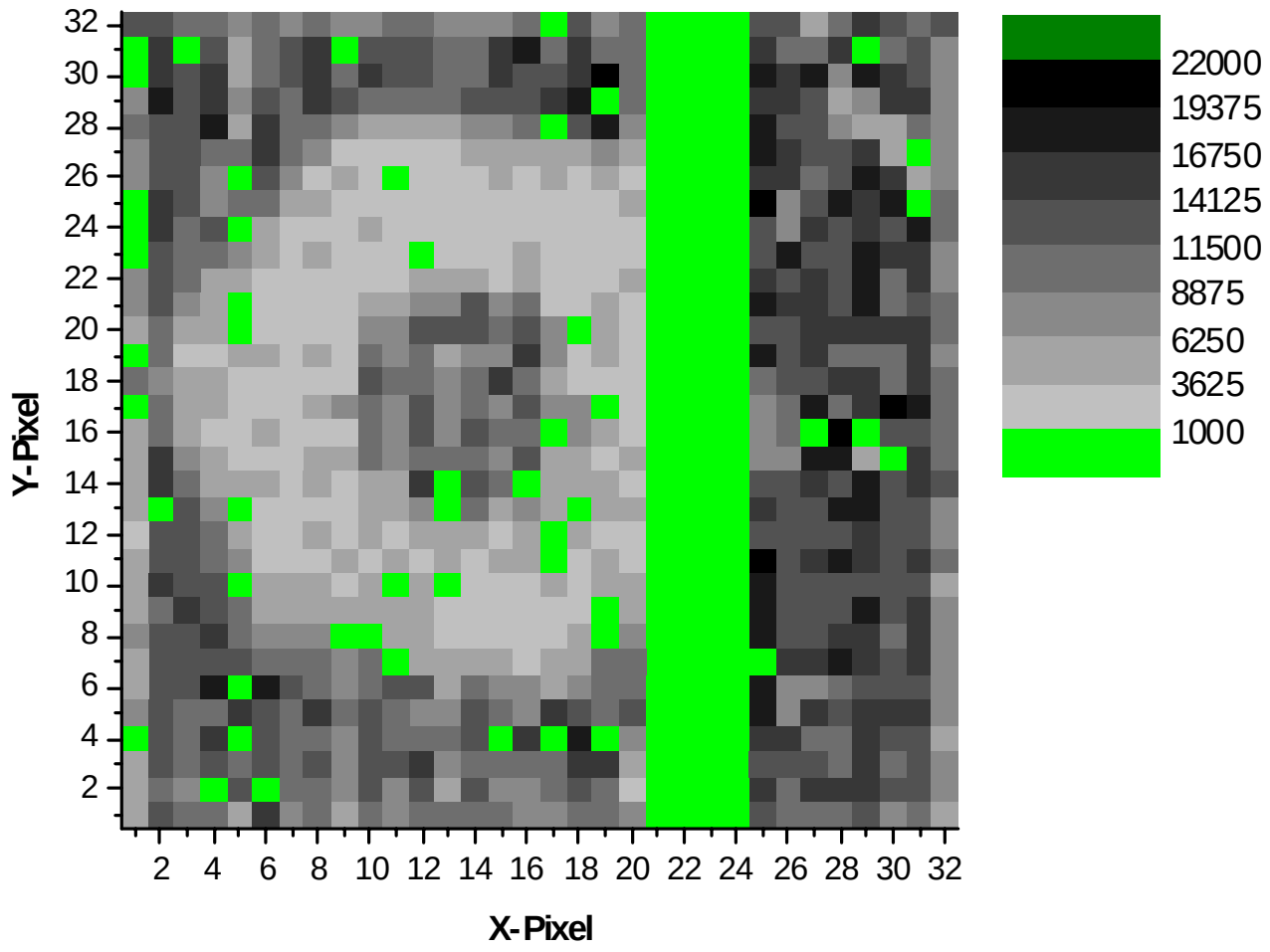

Fig. 9: Shadow image of a Nut, CdTe-Sensor exposed to $^{241}$Am 60 keV

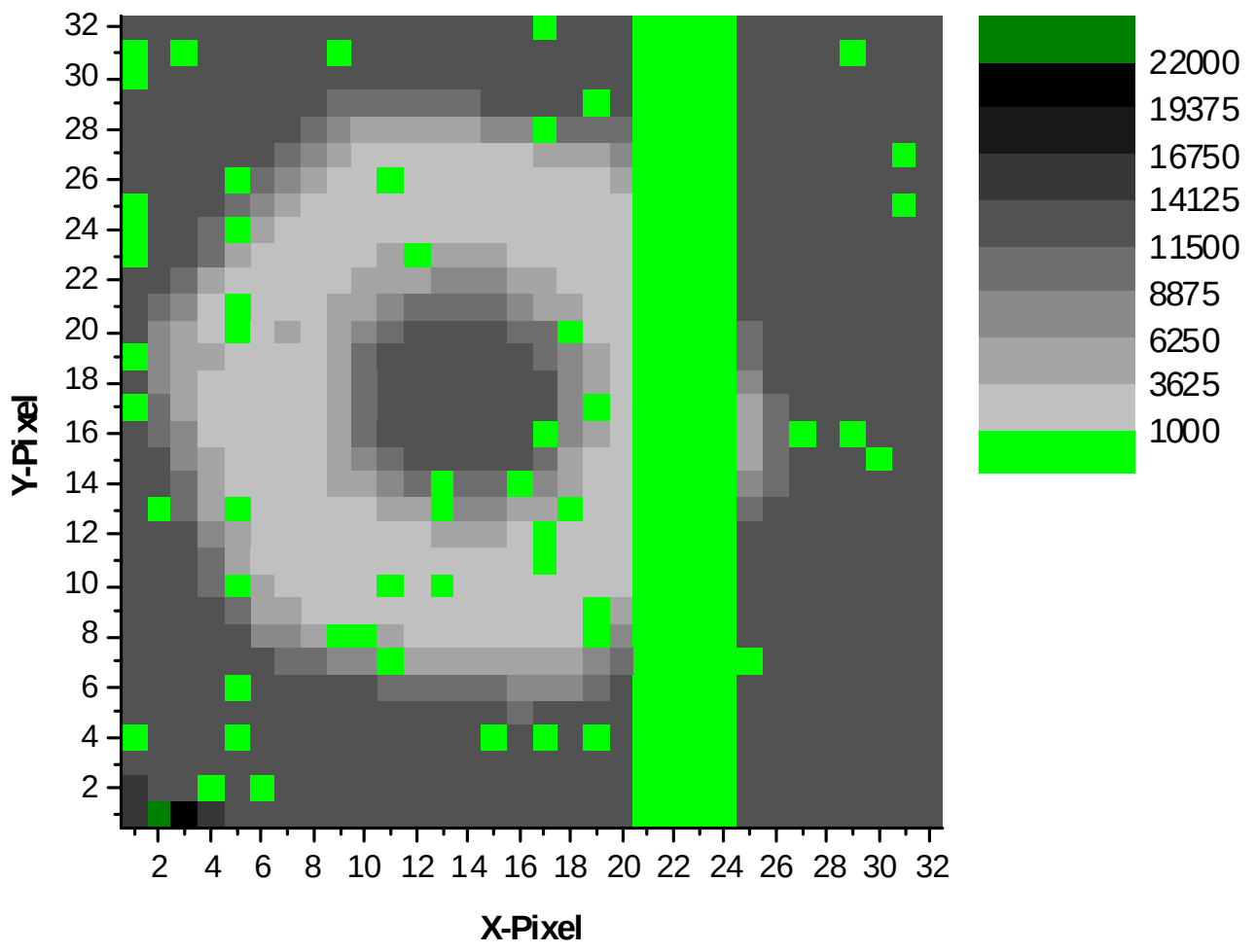

Fig. 10: Shadow image of a Nut, flat field corrected, CdTe-Sensor exposed to $^{241}$Am 60 keV

## Summary and conclusion

In this work new measurements with MPEC-chips are presented. The MPEC is a single photon counting readout chip with very low noise characteristics. The dispersion of the thresholds is very low and by means of two independent thresholds an energy window for incident photons can be selected. An excellent solder bump yield is achieved with a MPEC 2.1 chip attached to a Si sensor and the good homogeneity of the Si sensor could be demonstrated.

The newly designed MPEC 2.3 chip could successfully improve the performance of the MPEC 2.1 chip. A novel USB based data acquisition system has been developed serving as the read out system of the MPEC 2.3 detectors. In the near future a 2x2 detector module controlled by the USB system will be presented. A single MPEC 2.3 chip bump bonded onto a CdTe sensor was tested and could be operated successfully. Here, the gold stud bump bonding is still a crucial and challenging process. As CdTe suffers from inhomogeneous material properties, e.g. hole trapping, the image homogeneity is still poor but can be improved by a flat field correction.

The MPEC chips are designed for the AMS 0.8µm CMOS process offering only two metal layers. This makes effective shielding of the analog part from the digital part difficult and crosstalk via the sensor and the chip bulk is still present. This is a reason why the MPEC chips have to be operated with relatively high thresholds. The next chip generation will be designed in a 0.25µm-technology where up to five metal layers will make crosstalk reduction more effective.

## References

[1] P. Fischer et al., “A counting pixel readout chip for imaging applications”, Nucl. Instr. and Meth. A 405 (1998) 53-59

[2] M. Campbell et al., “Readout for a 64x64 Pixel Matrix with 15-bit Single Photon Counting” IEEE Trans. Nucl. Sci. Vol. 45 no.3 (1998) 751-753

[3] F. Anghinolfi et al., IEEE Trans. Nucl. Sci. NS-39 (1992) 654

[4] T. Takahashi, S. Watanabe: “Recent Progress in CdTe and CdZnTe Detectors”, submitted to IEEE Trans. Nucl. Sci. 2001

[5] T. Takahashi et al., “High Resolution CdTe Detector and Applications to Imaging Devices”, Talk given at IEEE Medical Imaging Conference Lyon, submitted to IEEE Trans. on Nucl. Sci., 2001

[6] P. Fischer et al., "A Counting CdTe Pixel Detector for hard X-ray and Gamma-ray Imaging", submitted to IEEE Trans. Nucl. Sci, 2001

[7] P. Fischer et al., “A Photon counting pixel chip with energy windowing", IEEE Trans. Nucl. Sci. vol. 47 no. 3 (2000) 881-884

[8] M. Lindner et al., "Comparison of hybrid pixel detectors with Si and GaAs sensors", Nucl. Instr. Meth. A 466 (2000) 63-73

[9] H. Krüger et al., “A X-Ray Imaging System using a Single Photon Counting CdTe Detector”, IEEE International Workshop on Room Temperature Semiconductor X- and Gamma-Ray Detectors, San Diego, USA, 2001

[10] M. Lindner et al., "Medical X-ray Imaging with Energy Windowing", Nucl. Instr. Meth. A 465 (2000) 229-234

[11] R. Irsigler et al. ”X-ray imaging using a 320*240 hybrid GaAs pixel detector”, IEEE Trans. Nucl. Sci., vol.46, no.3, June 1999, 507-512.

[12] C. Schwarz et al., "X-ray Imaging Using a Hybrid Photon Counting GaAs Pixel Detector", Nucl. Instr. Meth. B 78 (1999) 491-496

[13] National Institute of Standards and Technology, USA, 2001, “http://physics.nist.gov/PhysRefData/XrayMassCoef”